\documentclass[a4paper,10pt]{article}
\usepackage{amssymb}

\usepackage{amsmath}
\usepackage{fullpage}
\usepackage{boxedminipage}
\usepackage{listings}
\usepackage{minitoc}

\makeatletter \@addtoreset{equation}{section} \makeatother

\begin{document}


\begin{titlepage}

    \thispagestyle{empty}
    \begin{flushright}
        \hfill{CERN-PH-TH/2007-094} \\\hfill{UCLA/07/TEP/15}\\
    \end{flushright}

    \vspace{5pt}
    \begin{center}
        { \huge{\textbf{On the Moduli Space of non-BPS Attractors\\\vspace{5pt}for $\mathcal{N}=2$ Symmetric Manifolds}}}\vspace{25pt}
        \vspace{55pt}

        { \large{\textbf{{Sergio Ferrara$^{\diamondsuit\clubsuit\flat}$ and\ Alessio Marrani$^{\heartsuit\clubsuit}$}}}}\vspace{15pt}

        {$\diamondsuit$ \it Physics Department,Theory Unit, CERN, \\
        CH 1211, Geneva 23, Switzerland\\
        \texttt{sergio.ferrara@cern.ch}}

        \vspace{15pt}

        {$\clubsuit$ \it INFN - Laboratori Nazionali di Frascati, \\
        Via Enrico Fermi 40,00044 Frascati, Italy\\
        \texttt{marrani@lnf.infn.it}}

        \vspace{10pt}

         {$\flat$ \it Department of Physics and Astronomy,\\
        University of California, Los Angeles, CA USA\\
        \texttt{ferrara@physics.ucla.edu}}

         \vspace{10pt}

        {$\heartsuit$ \it Museo Storico della Fisica e\\
        Centro Studi e Ricerche ``Enrico Fermi"\\
        Via Panisperna 89A, 00184 Roma, Italy}

        \vspace{50pt}
\end{center}


\begin{abstract}We study the ``flat'' directions of non-BPS extremal black hole attractors
for $\mathcal{N}=2$, $d=4$ supergravities whose vector multiplets'
scalar manifold is endowed with homogeneous symmetric special
K\"{a}hler geometry. The non-BPS attractors with non-vanishing
central charge have a moduli space described by real special
geometry (and thus related to the $d=5$ parent theory), whereas the
moduli spaces of non-BPS attractors with vanishing central charge
are certain K\"{a}hler homogeneous symmetric manifolds. The
moduli spaces of the non-BPS attractors of the corresponding $\mathcal{N}=2$%
, $d=5$ theories are also indicated, and shown to be rank-$1$
homogeneous symmetric manifolds.
\end{abstract}

\end{titlepage}
\newpage

\section{Introduction\label{Intro}\protect\smallskip}

The issue of the Attractor Mechanism in extremal black holes \cite
{FKS,Strom,FK1,FK2,FGK} has recently received much attention, and a number
of interesting advances has been performed \cite{Sen-old1}-\nocite
{GIJT,Sen-old2,K1,TT,G,GJMT,Ebra1,K2,Ira1,Tom,BFM,AoB-book,FKlast,Ebra2,BFGM1,rotating-attr,K3,Misra1,Lust2,BFMY,CdWMa,DFT07-1,BFM-SIGRAV06,Cer-Dal,ADFT-2,Saraikin-Vafa-1,Ferrara-Marrani-1,TT2}
\cite{ADOT-1}. Among the others, we cite here the OSV conjecture \cite{OSV1}
(see also \cite{Saraikin-Vafa-1} and Refs. therein), relating black hole
(BH) entropy to topological partition functions,\ and the entropy function
formalism \cite{Sen-old1,Sen-old2}, which allows one to include the higher
derivative (gravitational and electromagnetic) corrections to
Maxwell-Einstein action (this is crucial specially for the so-called
``small'' BHs, with vanishing classical entropy). An important step has been
the realization that the Attractor Mechanism allows for extremal non-BPS BH
scalar configurations of different nature \cite{FGK,BFM,BFGM1} (see also
\cite{AoB-book}).

The present investigation concerns the latter issue, and in particular the
study of the ``flat'' directions of the Hessian\ matrix of the black hole
potential $V_{BH}$ at its critical points \cite
{ADFT,ADFFT,ADF2,ADF,TT,K3,TT2}. Beside considering the case of $\mathcal{N}%
=8$, $d=4,5$ supergravity, we will deal with $\mathcal{N}=2$, $d=4,5$
Maxwell-Einstein supergravity theories (which in the following treatment we
will simply call ``supergravities'') whose vector multiplets' scalar
manifold is homogeneous symmetric. Indeed, for such theories a rather
general analysis can be performed, determining the moduli space of the
various species of non-BPS critical points of $V_{BH}$, mainly by using
group theoretical methods (see \textit{e.g.} \cite{Helgason,Gilmore,Slansky}%
). In fact such moduli spaces are closely related to the nature (of the
stabilizer) of the ``orbits'' \cite{FG,FG2,Ferrara-Marrani-1} of the
background dyonic BH charge vector
\begin{equation}
\mathcal{Q}\equiv \left( m^{\Lambda },e_{\Lambda }\right)
\end{equation}
which supports the considered attractor, where $m^{\Lambda }$ and $%
e_{\Lambda }$ respectively stand for the magnetic and electric BH charges,
and $\Lambda =0,1,...,n_{V}$, with $n_{V}$ being the complex dimension of
the special K\"{a}hler scalar manifold. In the case of the $stu$ model \cite
{Duff-stu,BKRSW,K3}, our results are in agreement with the ones obtained in
\cite{BFGM1,Ferrara-Marrani-1,TT2}.\medskip

The paper is organized as follows.

In Sect. \ref{Sect2} we review the BPS and non-BPS critical points of $V_{BH,%
\mathcal{N}=2}$ for extremal BHs on homogeneous symmetric scalar manifolds,
and the corresponding orbits of the supporting BH charges \cite
{BFGM1,Ferrara-Marrani-1}. The resulting properties are summarized, in
particular the existence of ``flat'' directions for the non-BPS case,
related to the rank of the Hessian matrix of $V_{BH,\mathcal{N}=2}$ at the
corresponding critical points of $V_{BH,\mathcal{N}=2}$. Thence, in Sect.
\ref{Sect3} we deal with the $\mathcal{N}=8$ theory, and derive the moduli
spaces for non-singular\footnote{%
We will consider only \textit{non-singular} critical points of $V_{BH}$,
\textit{i.e.} solutions of the criticality conditions $\partial _{i}V_{BH}=0$
$\forall i$, such that $\left. V_{BH}\right| _{\partial V_{BH}=0}\neq 0$.} $%
\frac{1}{8}$-BPS and non-BPS critical points of $V_{BH,\mathcal{N}=8}$. In
Sect. \ref{Sect4} we do the same for the $\mathcal{N}=2$ supergravities
considered in Sect. \ref{Sect2}, by taking into account that in general
non-BPS critical points of $V_{BH,\mathcal{N}=2}$ can occur in two different
species, depending on the vanishing of the $\mathcal{N}=2$ central charge $Z$%
. Thus, in Sect. \ref{Sect5} we consider the case $d=5$, in particular the $%
\mathcal{N}=8$ theory (having only an $\frac{1}{8}$ non-singular class of
attractors) and the $\mathcal{N}=2$ homogeneous symmetric supergravities
(having an unique non-BPS class af attractors). Finally, some outlooks are
given in Sect. \ref{Conclusion}.

\section{$\mathcal{N}=2$, $d=4$ Homogeneous Symmetric Supergravities:\newline
Attractors and Critical Hessian \label{Sect2}}

The symmetric special K\"{a}hler manifolds $\frac{G_{V}}{H_{0}\otimes U(1)}$
of $\mathcal{N}=2$, $d=4$ supergravities have been classified in the
literature \cite{CVP,dWVVP}. \textbf{\ }With the exception of the family
whose prepotential is quadratic , all such theories can be obtained by
dimensional reduction of the $\mathcal{N}=2$, $d=5$ supergravities that were
constructed in \cite{GST1,GST2,GST3} (they will be treated in Sect. \ref
{Sect5}). The supergravities with symmetric manifolds that originate from $5$
dimensions all have cubic prepotentials determined by the norm form of the
Jordan algebra of degree $3$ that defines them \cite{GST1,GST2,GST3}.

The vector multiplets' scalar manifolds of homogeneous symmetric $\mathcal{N}%
=2$, $d=4$ supergravities are given in Table 1.

The irreducible sequence in the second row of Table 1 has quadratic
prepotentials (and thus $C_{ijk}=0$). On the other hand, the reducible
sequence in the third row, usually referred to as the \textit{generic Jordan
family}, has a $5$-dim. origin, and it is related tot the sequence $\mathbb{R%
}\oplus \mathbf{\Gamma }_{n}$ of reducible Euclidean Jordan algebras of
degree $3$. Here $\mathbb{R}$ denotes the $1$-dim. Jordan algebra and $%
\mathbf{\Gamma }_{n}$ denotes the Jordan algebra of degree $2$ associated
with a quadratic form of Lorentzian signature (see\footnote{%
In order to make contact with the notation used in the present paper, with
respect to the notation used in \cite{BFGM1} one has to shift $%
n+1\rightarrow n$ (and thus $n\in \mathbb{N}$) for the quadratic sequence,
and $n+2\rightarrow n$ (and thus $n\in \mathbb{N}$) for the cubic sequence.}
\textit{e.g.} Table 4 of \cite{BFGM1}, and Refs. therein).

\textbf{\ }Beside the generic Jordan family, there exist four other
supergravities defined by simple Jordan algebras of degree $3$. They are
called \textit{magic}, since their symmetry groups are the groups of the
famous \textit{Magic Square} of Freudenthal, Rozenfeld and Tits associated
with some remarkable geometries \cite{Freudenthal2,magic}. $J_{3}^{\mathbb{O}%
}$, $J_{3}^{\mathbb{H}}$, $J_{3}^{\mathbb{C}}$ and $J_{3}^{\mathbb{R}}$
denote the four simple Jordan algebras of degree $3$ with irreducible norm
forms, namely by the Jordan algebras of Hermitian $3\times 3$ matrices over
the four division algebras, \textit{i.e.} respectively over $\mathbb{A}=%
\mathbb{O}$ (octonions), $\mathbb{A}=\mathbb{H}$ (quaternions), $\mathbb{A}=%
\mathbb{C}$ (complex numbers) and $\mathbb{A}=\mathbb{R}$ (real numbers)
\cite{GST1,GST2,GST3,GST4,Jordan,Jacobson,Guna1,GPR}. By defining $A\equiv
dim_{\mathbb{R}}\mathbb{A}$ ($=8,4,2,1$ for $\mathbb{A}=\mathbb{O},\mathbb{H}%
,\mathbb{C},\mathbb{R}$, respectively), Tabel 1 yields that the complex
dimension of the scalar manifolds of the magic $\mathcal{N}=2$, $d=4$
supergravities is $3\left( A+1\right) $. Beside the analysis performed in
\cite{BFGM1}, Jordan algebras have been recently studied (and related to
extremal BHs) also in \cite{Ferrara-Gimon} and \cite{Rios}.

As found in \cite{FG}, the $\frac{1}{2}$-BPS supporting charge orbit is $%
\frac{G_{V}}{H_{0}}$. By denoting by $\widetilde{H}$ and $\widehat{H}$ two
non-compact forms of $H_{0}$, in \cite{BFGM1} it was found that the non-BPS $%
Z=0$ and non-BPS $Z\neq 0$ supporting BH charge orbits respectively are the
cosets $\frac{G_{V}}{\widetilde{H}}$ and $\frac{G_{V}}{\widehat{H}}$. Due to
the compact nature of $H_{0}$, the symmetry group of the $\frac{1}{2}$-BPS
critical points is the whole $H_{0}$, whereas the symmetry group of the
non-BPS $Z=0$ and non-BPS $Z\neq 0$ critical points respectively is the
\textit{maximal compact subgroup} (\textit{m.c.s.}) of $\widetilde{H}$ and $%
\widehat{H}$, in turn denoted by $\widetilde{h}$ and $\widehat{h}$
(actually, in the non-BPS $Z=0$ case, the symmetry is $\widetilde{h}^{\prime
}\equiv \frac{\widetilde{h}}{U(1)}$; see \cite{BFGM1} for further details).

The data of all the $\mathcal{N}=2$, $d=4$ homogeneous symmetric
supergravities are given in Tables 3 and 8 of \cite{BFGM1}.
\begin{table}[tbp]
\par
\begin{center}
\begin{tabular}{|c||c|c|c|}
\hline
& $
\begin{array}{c}
\\
\frac{G_{V}}{H_{V}} \\
~
\end{array}
$ & $
\begin{array}{c}
\\
r \\
~
\end{array}
$ & $
\begin{array}{c}
\\
dim_{\mathbb{C}}\equiv n_{V} \\
~
\end{array}
$ \\ \hline\hline
$
\begin{array}{c}
quadratic~sequence \\
n\in \mathbb{N}
\end{array}
$ & $\frac{SU(1,n)}{U(1)\otimes SU(n)}~$ & $1$ & $n~$ \\ \hline
$
\begin{array}{c}
\\
\mathbb{R}\oplus \mathbf{\Gamma }_{n},~n\in \mathbb{N} \\
~
\end{array}
$ & $\frac{SU(1,1)}{U(1)}\otimes \frac{SO(2,n)}{SO(2)\otimes SO(n)}~$ & $
\begin{array}{c}
\\
2~(n=1) \\
3~(n\geqslant 2)
\end{array}
~$ & $n+1$ \\ \hline
$
\begin{array}{c}
\\
J_{3}^{\mathbb{O}} \\
~
\end{array}
$ & $\frac{E_{7(-25)}}{E_{6(-78)}\otimes U(1)}$ & $3$ & $27$ \\ \hline
$
\begin{array}{c}
\\
J_{3}^{\mathbb{H}} \\
~
\end{array}
$ & $\frac{SO^{\ast }(12)}{U(6)}~$ & $3$ & $15$ \\ \hline
$
\begin{array}{c}
\\
J_{3}^{\mathbb{C}} \\
~
\end{array}
$ & $\frac{SU(3,3)}{S\left( U(3)\otimes U(3)\right) }=\frac{SU(3,3)}{%
SU(3)\otimes SU(3)\otimes U(1)}$ & $3$ & $9~$ \\ \hline
$
\begin{array}{c}
\\
J_{3}^{\mathbb{R}} \\
~
\end{array}
$ & $\frac{Sp(6,\mathbb{R})}{U(3)}$ & $3$ & $6$ \\ \hline
\end{tabular}
\end{center}
\caption{$\mathcal{N}$\textbf{$=2$, $d=4$ homogeneous symmetric special
K\"{a}hler manifolds}}
\end{table}

In the following treatment we will denote by $\frak{r}$ the rank of the $%
2n_{V}\times 2n_{V}$ Hessian matrix $\mathbf{H}$ of $V_{BH}$. Since in $%
\mathcal{N}=2$, $d=4$ supergravity the $\frac{1}{2}$-BPS critical points of $%
V_{BH}$ are stable, and $\mathbf{H}_{\frac{1}{2}-BPS}$ has no massless modes
\cite{FGK}, it holds that the rank is maximal: $\frak{r}_{\frac{1}{2}%
-BPS}=2n_{V}$. On the other hand, from the analysis performed in \cite{BFGM1}
for homogeneous symmetric $\mathcal{N}=2$, $d=4$ supergravities, it follows
that $\frak{r}_{non-BPS}$ is model-dependent, and it also depends on the
vanishing of the $\mathcal{N}=2$ central charge $Z$.

In the quadratic sequence $\frac{SU(1,n)}{U(1)\otimes SU(n)}$ ($n\in \mathbb{%
N}$), only non-BPS critical points with $Z=0$ exist. In this case, $\frak{r}%
_{non-BPS,Z=0}=2$, and $\mathbf{H}_{non-BPS,Z=0}$ has $2n-2=2n_{V}-2$
massless modes.

For the generic Jordan family $\frac{SU(1,1)}{U(1)}\otimes \frac{SO(2,n)}{%
SO(2)\otimes SO(n)}$ ($n\in \mathbb{N}$), it holds that $\frak{r}%
_{non-BPS,Z\neq 0}=n+2=n_{V}+1$ ($\mathbf{H}_{non-BPS,Z\neq 0}$ has $%
n=n_{V}-1$ massless modes), whereas $\frak{r}_{non-BPS,Z=0}=6$ ($\mathbf{H}%
_{non-BPS,Z=0}$ has $2n-4=2n_{V}-6$ massless modes).

Concerning the magic models, it holds that $\frak{r}_{non-BPS,Z\neq
0}=3A+4=n_{V}+1$ ($\mathbf{H}_{non-BPS,Z\neq 0}$ has $3A+2=n_{V}-1$ massless
modes), whereas $\frak{r}_{non-BPS,Z=0}=2A+6$ ($\mathbf{H}_{non-BPS,Z=0}$
has $4A$ massless modes).

Thus, the above findings match the result found by Tripathy and Trivedi in
\cite{TT} for a generic special K\"{a}hler $d$-geometry\footnote{%
Following the notation of \cite{dWVVP}, by $d$-geometry we mean a special K%
\"{a}hler geometry based on an holomorphic prepotential function of the
cubic form $F\left( X\right) =d_{ABC}\frac{X^{A}X^{B}X^{C}}{X^{0}}$ ($A$, $B$%
, $C=0,1,...,n_{V}$).} of complex dimension $n_{V}$: $\frak{r}%
_{non-BPS,Z\neq 0}=n_{V}+1$, \textit{i.e. }$\mathbf{H}_{non-BPS,Z\neq 0}$
has $n_{V}-1$ massless modes.

\section{$\mathcal{N}=8$, $d=4$ Supergravity:\newline
Attractors and their Moduli Spaces\label{Sect3}}

In order to understand the moduli spaces of the two classes ($Z\neq 0$ and $%
Z=0$) of non-BPS attractors of homogeneous symmetric $\mathcal{N}=2$, $d=4$
supergravities, it is instructive to consider $\mathcal{N}=8$, $d=4$
supergravity, based on the real $70$-dim. homogeneous symmetric manifold $%
\frac{G_{8}}{H_{8}}=\frac{E_{7(7)}}{SU(8)}$.

From the analysis performed in \cite{FG,FM,FKlast} it holds that only two
non-singular classes of critical points of $V_{BH,\mathcal{N}=8}$ exist (see
also \cite{Ferrara-Marrani-1}): the $\frac{1}{8}$-BPS class, supported by
the BH charge orbit $\mathcal{O}_{\frac{1}{8}-BPS,\mathcal{N}=8}\equiv \frac{%
G_{8}}{\mathcal{H}_{0}}=\frac{E_{7(7)}}{E_{6(2)}}$, and the non-BPS class,
supported by the BH charge orbit $\mathcal{O}_{non-BPS,\mathcal{N}=8}\equiv
\frac{G_{8}}{\widehat{\mathcal{H}}_{0}}=\frac{E_{7(7)}}{E_{6(6)}}$. Thus,
the $\frac{1}{8}$-BPS and non-BPS orbits respectively correspond to the
maximal (non-compact) subgroup of $E_{7(7)}$ to be $E_{6(2)}\otimes U(1)$
and $E_{6(6)}\otimes SO(1,1)$, where $E_{6(2)}$ and $E_{6(6)}$ are two
non-compact forms of the exceptional group $E_{6}\equiv E_{6(-78)}$ \cite
{Gilmore}. The $70\times 70$ $\frac{1}{8}$-BPS Hessian $\mathbf{H}_{\frac{1}{%
8}-BPS,\mathcal{N}=8}$ has rank $30$, with $40$ massless modes \cite{ADF2}
sitting in the representation $\left( \mathbf{20},\mathbf{2}\right) $ of the
enhanced $\frac{1}{8}$-BPS symmetry group $SU(6)\otimes SU(2)=m.c.s.\left(
\mathcal{H}_{0}\right) $ \cite{Ferrara-Marrani-1}. On the other hand, the $%
70\times 70$ non-BPS Hessian $\mathbf{H}_{non-BPS,\mathcal{N}=8}$ has rank $%
28$, with $42$ massless modes sitting in the representation $\mathbf{42}$ of
the enhanced non-BPS symmetry group $USp(8)=m.c.s.\left( \widehat{\mathcal{H}%
}_{0}\right) $ \cite{Ferrara-Marrani-1}.

As it will be evident from the reasoning performed below, the massless modes
of the Hessian of $V_{BH,\mathcal{N}=8}$ at its non-singular $\frac{1}{8}$%
-BPS\ and non-BPS critical points actually are ``flat'' directions of $V_{BH,%
\mathcal{N}=8}$ at the corresponding critical points. Such ``flat''
directions span the following real homogeneous symmetric sub-manifolds of $%
\frac{E_{7(7)}}{SU(8)}$:
\begin{equation}
\begin{array}{l}
\frac{1}{8}-BPS~\text{\textit{moduli~space}}:\frac{\mathcal{H}_{0}}{%
m.c.s.\left( \mathcal{H}_{0}\right) }=\frac{E_{6(2)}}{SU(6)\otimes SU(2)}%
,~dim_{\mathbb{R}}=40; \\
\\
non-BPS~\text{\textit{moduli~space}}:\frac{\widehat{\mathcal{H}}_{0}}{%
m.c.s.\left( \widehat{\mathcal{H}}_{0}\right) }=\frac{E_{6(6)}}{USp(8)}%
,~dim_{\mathbb{R}}=42.
\end{array}
\label{N=8-moduli-spaces}
\end{equation}

Both moduli spaces $\frac{E_{6(2)}}{SU(6)\otimes SU(2)}$ and $\frac{E_{6(6)}%
}{USp(8)}$ share the same structure: they are the coset of the (non-compact)
stabilizer of the corresponding supporting BH charge orbit and of its
m.c.s.. As yielded by the analysis performed in Sect. \ref{Sect4}, this is
also the structure of the moduli spaces of the two classes of non-BPS
attractors of the homogeneous symmetric $\mathcal{N}=2$, $d=4$
supergravities.

Remarkably, $\frac{E_{6(6)}}{USp(8)}$ is the real manifold on which $%
\mathcal{N}=8$, $d=5$ supergravity is based. Such a relation with the $d=5$
parent theory is exhibited also by non-BPS $Z\neq 0$ moduli spaces of the
homogeneous symmetric $\mathcal{N}=2$, $d=4$ supergravities; see Sect. \ref
{Sect4}.

In order to understand that the ``flat'' directions of the Hessian of
\begin{equation}
V_{BH,\mathcal{N}=8}\equiv \frac{1}{2}Z_{AB}\left( \phi ,Q\right) \overline{Z%
}^{AB}\left( \phi ,Q\right)
\end{equation}
at its critical points actually span a moduli space, it is useful to recall
that the $\mathcal{N}=8$ central charge matrix $Z_{AB}\left( \phi ,Q\right) $
can be rewritten as \cite{DFL}
\begin{equation}
Z_{AB}\left( \phi ,Q\right) =\left( Q^{T}L\left( \phi \right) \right)
_{AB}=\left( Q^{T}\right) _{\Lambda }L_{AB}^{\Lambda }\left( \phi \right) ,
\label{central-charge-parameterization}
\end{equation}
where $\phi $ denote the $70$ real scalar fields parameterizing the coset $%
\frac{G_{8}}{H_{8}}=\frac{E_{7(7)}}{SU(8)}$, $Q$ is the $\mathcal{N}=8$
charge vector, and $L_{AB}^{\Lambda }\left( \phi \right) \in G_{8}$ is the
field-dependent coset representative, \textit{i.e.} a local section of the
principal bundle $G_{8}$ over $\frac{G_{8}}{H_{8}}$ with structure group $%
H_{8}$. Thus, it follows that
\begin{equation}
V_{BH,\mathcal{N}=8}\left( \phi ,Q\right) =V_{BH,\mathcal{N}=8}\left( \phi
_{g},Q^{g}\right) =V_{BH,\mathcal{N}=8}\left( \phi _{g},\left( g^{-1}\right)
^{T}Q\right) ,
\end{equation}
which shows that $V_{BH,\mathcal{N}=8}$ is not $G_{8}$-invariant, because
its coefficients (given by the components of $Q$) do not in general remain
the same.

Now, if we take $g\equiv g_{Q}\in H_{Q}$, where $H_{Q}$ is the stabilizer of
one of the orbits $\frac{G_{8}}{H_{Q}}$ spanned by the charge vector $Q$,
then $Q^{g_{Q}}=Q$, and thus:
\begin{equation}
V_{BH,\mathcal{N}=8}\left( \phi ,Q\right) =V_{BH,\mathcal{N}=8}\left( \phi
_{g_{Q}},Q\right) .  \label{11june}
\end{equation}
Let us now split the fields $\phi $ into $\phi _{Q}\in \frac{H_{Q}}{h_{Q}}$
(where $h_{Q}\equiv m.c.s.\left( H_{Q}\right) $) and into the remaining $%
\widehat{\phi }_{Q}$, paremeterizing the complement of $\frac{H_{Q}}{h_{Q}}$
in $\frac{G_{8}}{H_{Q}}$. By defining
\begin{equation}
V_{BH,\mathcal{N}=8,crit}\left( \phi _{Q},Q\right) \equiv \left. V_{BH,%
\mathcal{N}=8}\left( \phi ,Q\right) \right| _{\frac{\partial V_{BH,\mathcal{N%
}=8}}{\partial \widehat{\phi }_{Q}}=0},
\end{equation}
Eq. (\ref{11june}) yields the invariance of $V_{BH,\mathcal{N}=8,crit}\left(
\phi _{Q},Q\right) $ under $H_{Q}$:
\begin{equation}
V_{BH,\mathcal{N}=8,crit}\left( \left( \phi _{Q}\right) _{g_{Q}},Q\right)
=V_{BH,\mathcal{N}=8,crit}\left( \phi _{Q},Q\right) .
\end{equation}
Since $H_{Q}$ is a non-compact group, this implies $V_{BH,\mathcal{N}=8}$ to
be independent \textit{at its critical points} on the fields $\phi _{Q}$
parameterizing the coset $\frac{H_{Q}}{h_{Q}}$. In other words, the
(covariant) derivatives of $V_{BH,\mathcal{N}=8}$, when evaluated \textit{at
its critical points} and with all indices spanning the coset $\frac{H_{Q}}{%
h_{Q}}$, vanish at \textit{all} orders.

It is easy to realize that such a reasoning can be performed for all
supergravities with $\mathcal{N}\geqslant 1$ based on homogeneous (not
necessarily symmetric) manifolds\footnote{%
This is actually always the case for $\mathcal{N}\geqslant 3$ (see \textit{%
e.g.} \cite{ADFT}).} $\frac{G_{\mathcal{N}}}{H_{\mathcal{N}}}$, also in
presence of matter multiplets (and thus of matter charges). Indeed, such
arguments also apply to a generic, not necessarily supersymmetric,
Maxwell-Einstein system with an homogeneous (not necessarily symmetric)
scalar manifold.

By choosing $Q$ belonging to an orbit of the representation $R_{V}$ of $G_{%
\mathcal{N}}$ which supports critical points of $V_{BH,\mathcal{N}}$, the
previous reasoning yields the interesting result that, \textit{up to
``flat'' directions (at all orders in covariant differentiation of }$V_{BH,%
\mathcal{N}}$\textit{)}, \textit{all} critical points of $V_{BH,\mathcal{N}}$
in \textit{all} $\mathcal{N}\geqslant 0$ Maxwell-Einstein (super)gravities
with an homogeneous (not necessarily symmetric) scalar manifold (also in
presence of matter multiplets) are \textit{stable}, and thus they are
\textit{attractors} in a generalized sense.

\section{$\mathcal{N}=2$, $d=4$ Symmetric Supergravities:\newline
Attractors and their Moduli Spaces\label{Sect4}}

By using the arguments of the previous Section, we now determine the moduli
spaces of non-BPS critical points of $V_{BH,\mathcal{N}=2}$ (with $Z\neq 0$
and $Z=0$) for all $\mathcal{N}=2$, $d=4$ homogeneous symmetric
supergravities.

As previously noticed, $\mathcal{N}=2$ $\frac{1}{2}$-BPS critical points are
\textit{stable}, and at such points all the scalars are stabilized by the
classical Attractor Mechanism, because $\mathbf{H}_{\frac{1}{2}-BPS}$ has no
massless modes at all \cite{FGK}; thus, there is no $\frac{1}{2}$-BPS moduli
space for all $\mathcal{N}=2$, $d=4$ supergravities (as far as the metric of
the scalar manifold is non-singular and positive-definite). This is
qualitatively different from the previously considered case of $\mathcal{N}=8
$ $\frac{1}{8}$-BPS critical points.

In the framework of $\mathcal{N}=2$, $d=4$ homogeneous symmetric
supergravities, such a difference can be traced back to the fact that the
stabilizer of the $\mathcal{N}=2$ charge orbit $\mathcal{O}_{\frac{1}{2}-BPS,%
\mathcal{N}=2}$ is \textit{compact} (see Tables 3 and 8 of \cite{BFGM1}).

In general, such a difference can be explained by noticing that for $%
\mathcal{N}=2$ the $\frac{1}{\mathcal{N}}=\frac{1}{2}$-BPS configurations
are the \textit{maximally supersymmetric} ones, \textit{i.e.} they preserve
the maximum number of supersymmetries out of the ones related to the
asymptotically flat BH background. For $2<\mathcal{N}\leqslant 8$ the $\frac{%
1}{\mathcal{N}}$-BPS configurations are \textit{not maximally supersymmetric}%
, and the configurations preserving the maximum number of supersymmetries
have vanishing classical BH entropy.

It is now possible to determine the moduli spaces of non-BPS critical points
of $V_{BH,\mathcal{N}=2}$ (with $Z\neq 0$ and $Z=0$) for all $\mathcal{N}=2$%
, $d=4$ homogeneous symmetric supergravities (which match the results about
the rank of the Hessian reported in Sect. \ref{Sect2}). Consistently with
the notation introduced in Sect. \ref{Sect2}, the $\mathcal{N}=2$ non-BPS $%
Z\neq 0$ moduli space is the coset $\frac{\widehat{H}}{\widehat{h}}$,
whereas the $\mathcal{N}=2$ non-BPS $Z=0$ moduli space is the coset $\frac{%
\widetilde{H}}{\widetilde{h}}=\frac{\widetilde{H}}{\widetilde{h}^{\prime
}\otimes U(1)}$ (see \cite{BFGM1} for further details on notation). They are
respectively given by Table 2 and 3.

\begin{table}[tbp]
\par
\begin{center}
\begin{tabular}{|c||c|c|c|}
\hline
& $
\begin{array}{c}
\\
\frac{\widehat{H}}{\widehat{h}} \\
~
\end{array}
$ & $
\begin{array}{c}
\\
r \\
~
\end{array}
$ & $
\begin{array}{c}
\\
dim_{\mathbb{R}} \\
~
\end{array}
$ \\ \hline
$
\begin{array}{c}
\\
\mathbb{R}\oplus \mathbf{\Gamma }_{n},~n\in \mathbb{N} \\
~
\end{array}
$ & $SO(1,1)\otimes \frac{SO(1,n-1)}{SO(n-1)}~$ & $
\begin{array}{c}
\\
1~(n=1) \\
2~(n\geqslant 2)
\end{array}
~$ & $n$ \\ \hline
$
\begin{array}{c}
\\
J_{3}^{\mathbb{O}} \\
~
\end{array}
$ & $\frac{E_{6(-26)}}{F_{4(-52)}}$ & $2$ & $26$ \\ \hline
$
\begin{array}{c}
\\
J_{3}^{\mathbb{H}} \\
~
\end{array}
$ & $\frac{SU^{\ast }(6)}{USp(6)}~$ & $2$ & $14$ \\ \hline
$
\begin{array}{c}
\\
J_{3}^{\mathbb{C}} \\
~
\end{array}
$ & $\frac{SL(3,\mathbb{C})}{SU(3)}$ & $2$ & $8~$ \\ \hline
$
\begin{array}{c}
\\
J_{3}^{\mathbb{R}} \\
~
\end{array}
$ & $\frac{SL(3,\mathbb{R})}{SO(3)}$ & $2$ & $5$ \\ \hline
\end{tabular}
\end{center}
\caption{\textbf{Moduli spaces of non-BPS }$Z\neq 0$\textbf{\ critical
points of }$V_{BH,\mathcal{N}=2}$ \textbf{in }$\mathcal{N}$\textbf{$=2$, $%
d=4 $ homogeneous symmetric supergravities. They are the }$\mathcal{N}$$=2$%
\textbf{,} $d=5$ \textbf{homogeneous symmetric real special manifolds}}
\end{table}
\begin{table}[tbp]
\par
\begin{center}
\begin{tabular}{|c||c|c|c|}
\hline
& $
\begin{array}{c}
\\
\frac{\widetilde{H}}{\widetilde{h}}=\frac{\widetilde{H}}{\widetilde{h}%
^{\prime }\otimes U(1)} \\
~
\end{array}
$ & $
\begin{array}{c}
\\
r \\
~
\end{array}
$ & $
\begin{array}{c}
\\
dim_{\mathbb{C}} \\
~
\end{array}
$ \\ \hline\hline
$
\begin{array}{c}
quadratic~~sequence \\
n\in \mathbb{N}
\end{array}
$ & $\frac{SU(1,n-1)}{U(1)\otimes SU(n-1)}~$ & $1$ & $n~-1$ \\ \hline
$
\begin{array}{c}
\\
\mathbb{R}\oplus \mathbf{\Gamma }_{n},~n\in \mathbb{N} \\
~
\end{array}
$ & $\frac{SO(2,n-2)}{SO(2)\otimes SO(n-2)},~n\geqslant 3~$ & $
\begin{array}{c}
\\
1~(n=3) \\
2~(n\geqslant 4)
\end{array}
~$ & $n-2$ \\ \hline
$
\begin{array}{c}
\\
J_{3}^{\mathbb{O}} \\
~
\end{array}
$ & $\frac{E_{6(-14)}}{SO(10)\otimes U(1)}$ & $2$ & $16$ \\ \hline
$
\begin{array}{c}
\\
J_{3}^{\mathbb{H}} \\
~
\end{array}
$ & $\frac{SU(4,2)}{SU(4)\otimes SU(2)\otimes U(1)}~$ & $2$ & $8$ \\ \hline
$
\begin{array}{c}
\\
J_{3}^{\mathbb{C}} \\
~
\end{array}
$ & $\frac{SU(2,1)}{SU(2)\otimes U(1)}\otimes \frac{SU(1,2)}{SU(2)\otimes
U(1)}$ & $2$ & $4$ \\ \hline
$
\begin{array}{c}
\\
J_{3}^{\mathbb{R}} \\
~
\end{array}
$ & $\frac{SU(2,1)}{SU(2)\otimes U(1)}$ & $1$ & $2$ \\ \hline
\end{tabular}
\end{center}
\caption{\textbf{Moduli spaces of non-BPS }$Z=0$\textbf{\ critical points of
}$V_{BH,\mathcal{N}=2}$ \textbf{in }$\mathcal{N}$\textbf{$=2$, $d=4$
homogeneous symmetric supergravities. They are (non-special) homogeneous
symmetric K\"{a}hler manifolds}}
\end{table}

Remarkably, the moduli spaces of non-BPS $Z\neq 0$ critical points are
nothing but the $\mathcal{N}=2$, $d=5$ homogeneous symmetric real special
manifolds, \textit{i.e.} the scalar manifolds of the $d=5$ parents of the
considered theories. Their real dimension $dim_{\mathbb{R}}$ (rank $r$) is
the complex dimension $dim_{\mathbb{C}}$ (rank $r$) of the $\mathcal{N}=2$, $%
d=4$ symmetric special K\"{a}hler manifolds listed in Table 1, minus one.
With the exception of the $st^{2}$ model ($n=1$ element of the generic
Jordan family) having $\frac{\widehat{H}}{\widehat{h}}=SO(1,1)$ with rank $%
r=1$, all such moduli spaces have rank $r=2$. The results of Table 2 are
consistent with the non-BPS $Z\neq 0$ ``$n_{V}+1$ / $n_{V}-1$'' mass
degeneracy splitting found by Tripathy and Trivedi in \cite{TT} (and
confirmed in \cite{BFGM1,TT2,Ferrara-Marrani-1}) for a generic special
K\"{a}hler $d$-geometry of complex dimension $n_{V}$.

Concerning the moduli spaces of non-BPS $Z=0$ critical points, they are
homogeneous symmetric (not special) K\"{a}hler manifolds. In the models $%
st^{2}$ and $stu$ ($n=1$ and $n=2$ elements of the generic Jordan family)
there are no non-BPS $Z=0$ ``flat'' directions at all (see Appendix II of
\cite{BFGM1}). By recalling that $A\equiv dim_{\mathbb{R}}\mathbb{A}$, Tabel
3 yields that the the moduli spaces of non-BPS $Z=0$ critical points of $%
V_{BH,\mathcal{N}=2}$ in magic $\mathcal{N}=2$, $d=4$ supergravities have
complex dimension $2A$. Interestingly, for the $\mathcal{N}=2$, $d=4$ magic
supergravity associated to $J_{3}^{\mathbb{O}}$, the non-BPS $Z=0$ moduli
space is the manifold $\frac{E_{6(-14)}}{SO(10)\otimes U(1)}$, which is
related to another exceptional Jordan triple system over $\mathbb{O}$, as
found long time ago by G\"{u}naydin, Sierra and Townsend \cite{GST1,GST2}.

As mentioned in the Introduction, all this is consistent with the results
about the $stu$ model \cite{Duff-stu,BKRSW,K3} obtained in \cite
{BFGM1,Ferrara-Marrani-1,TT2}: for such a model ($n=2$ element of the
generic Jordan family) there are $2$ non-BPS $Z\neq 0$ ``flat'' directions
(spanning the manifold $\left( SO(1,1)\right) ^{2}$, as yielded by Table 2)
and no non-BPS $Z=0$ ``flat'' directions.

\section{$d=5$, $\mathcal{N}=8$ and $\mathcal{N}=2$ Symmetric Supergravities:%
\newline
Attractors and their Moduli Spaces\label{Sect5}}

$\mathcal{N}=8$, $d=5$ supergravity, based on the homogeneous symmetric real
manifold $\frac{E_{6(6)}}{USp(8)}$ ($dim_{\mathbb{R}}=42$), has only one
non-singular (\textit{i.e. }with non-vanishing cubic invariant $I_{3}$)
charge orbit, namely the $\frac{1}{8}$-BPS one \cite{FM,FG,FG2}:
\begin{equation}
\frac{E_{6(6)}}{F_{4(4)}}.
\end{equation}
The $d=5$ supersymmetry reduction $\mathcal{N}=8\longrightarrow \mathcal{N}=2
$ gives $14$ vector multiplets and $7$ hypermultiplets \cite{ADF-d=5}
corresponding to the two ``extremal'' (in the sense of having the maximum
number of vector multiplets or hypermultiplets) truncations \cite{ADF-d=5}:
\begin{equation}
\begin{array}{l}
\left( n_{V},n_{H}\right) =\left( 14,0\right) :\frac{SU^{\ast }(6)}{USp(6)}~%
\text{\textit{real special}}; \\
\\
\left( n_{V},n_{H}\right) =\left( 0,7\right) :\frac{F_{4(4)}}{USp(6)\otimes
USp(2)}~\text{\textit{quaternionic K\"{a}hler}},
\end{array}
\end{equation}
yielding $14$ massive and $28$ massless modes of $\mathbf{H}_{\frac{1}{8}%
-BPS,\mathcal{N}=8,d=5}$. Thus, the moduli space of the non-singular $\frac{1%
}{8}$-BPS critical points of $V_{BH,\mathcal{N}=8}$ in $\mathcal{N}=8,d=5$
supergravity is given by the quaternionic K\"{a}hler manifold
\begin{equation}
\frac{F_{4(4)}}{USp(6)\otimes USp(2)}.
\end{equation}

Considering now the case $\mathcal{N}=2$, the manifolds\textbf{\ }of the
homogeneous symmetric $\mathcal{N}=2$, $d=5$ supergravities are given by
Table 2. As shown in \cite{FG2}, the $\frac{1}{2}$-BPS critical points are
stable already at the Hessian level, as in the $d=4$ case. There is an
unique class of non-singular non-BPS critical points; by slightly modifying
the notation introduced in \cite{FG2}, we denote by $\widetilde{H}_{5}$ and $%
\widetilde{K}_{5}$ the (non-compact) stabilizer of the corresponding non-BPS
charge orbits and its m.c.s., respectively. It then follows that the moduli
space of the unique class of non-singular non-BPS critical points of $V_{BH,%
\mathcal{N}=2}$ in homogeneous symmetric $\mathcal{N}=2$, $d=5$
supergravities is given by the homogeneous symmetric manifold
\begin{equation}
\frac{\widetilde{H}_{5}}{\widetilde{K}_{5}}.
\end{equation}
The explicit form of $\frac{\widetilde{H}_{5}}{\widetilde{K}_{5}}$ and its
data for all homogeneous symmetric $\mathcal{N}=2$, $d=5$ supergravities is
given in Table 4. Such a Table yields that the the moduli spaces of
non-singular non-BPS critical points of $V_{BH,\mathcal{N}=2}$ in magic $%
\mathcal{N}=2$, $d=5$ supergravities have real dimension $2A$. Their
stabilizer contains the group $spin\left( 1+A\right) $. Here we just point
out that, unlike the case $d=4$ \cite{TT,TT2}, an explicit calculation of
the ``flat'' directions of non-BPS critical points of $V_{BH,\mathcal{N}=2}$
in $d=5$, despite some recent works on Attractor Mechanism and entropy
function formalism in $d=5$ supergravities (see \textit{e.g.} \cite
{Larsen-rev}, \cite{Cardoso}\ and \cite{Goldstein}, and Refs. therein), is
missing at the present time.

\begin{table}[tbp]
\par
\begin{center}
\begin{tabular}{|c||c|c|c|}
\hline
& $
\begin{array}{c}
\\
\frac{\widetilde{H}_{5}}{\widetilde{K}_{5}} \\
~
\end{array}
$ & $
\begin{array}{c}
\\
r \\
~
\end{array}
$ & $
\begin{array}{c}
\\
dim_{\mathbb{R}} \\
~
\end{array}
$ \\ \hline
$
\begin{array}{c}
\\
\mathbb{R}\oplus \mathbf{\Gamma }_{n},~n\in \mathbb{N} \\
~
\end{array}
$ & $\frac{SO(1,n-2)}{SO(n-2)},~n\geqslant 3~$ & $
\begin{array}{c}
\\
1~(n\geqslant 3) \\
~
\end{array}
~$ & $n-2$ \\ \hline
$
\begin{array}{c}
\\
J_{3}^{\mathbb{O}} \\
~
\end{array}
$ & $\frac{F_{4(-20)}}{SO(9)}$ & $1$ & $16$ \\ \hline
$
\begin{array}{c}
\\
J_{3}^{\mathbb{H}} \\
~
\end{array}
$ & $\frac{USp(4,2)}{USp(4)\otimes USp(2)}~$ & $1$ & $8$ \\ \hline
$
\begin{array}{c}
\\
J_{3}^{\mathbb{C}} \\
~
\end{array}
$ & $\frac{SU(2,1)}{SU(2)\otimes U(1)}$ & $1$ & $4~$ \\ \hline
$
\begin{array}{c}
\\
J_{3}^{\mathbb{R}} \\
~
\end{array}
$ & $\frac{SL(2,\mathbb{R})}{SO(2)}$ & $1$ & $2$ \\ \hline
\end{tabular}
\end{center}
\caption{\textbf{Moduli spaces of non-BPS critical points of }$V_{BH,%
\mathcal{N}=2}$ \textbf{in }$\mathcal{N}$\textbf{$=2$, $d=5$ homogeneous
symmetric supergravities}}
\end{table}

\section{\label{Conclusion}\textbf{Conclusion}}

In the present investigation we have extended the analysis performed in \cite
{BFGM1} and \cite{Ferrara-Marrani-1} about the spectrum of non-BPS critical
points of $V_{BH,\mathcal{N}=2}$, their degeneracy and stability. For the
case of $d$-geometries \cite{TT,TT2}, and in particular for homogeneous
symmetric special K\"{a}hler geometries \cite{BFGM1,Ferrara-Marrani-1}, the
Hessian matrix of $V_{BH,\mathcal{N}=2}$ at its non-BPS critical points
generally has some strictly positive eigenvalues and some vanishing
eigenvalues, corresponding to ``flat'' directions. For the non-BPS $Z\neq 0$
case, our analysis generalizes the findings of \cite{TT2}.

One should not be surprised by our result, because the existence of ``flat''
directions in the Hessian of $V_{BH}$ was pointed out also at BPS critical
points (preserving $4$ supersymmetries) in the framework of $\mathcal{N}>2$,
$d=4$ extended supergravities \cite{ADF2,Ferrara-Marrani-1}, the ``flat''
directions being associated to hypermultiplets' scalar degrees of freedom in
the supersymmetry reduction $\mathcal{N}>2\longrightarrow \mathcal{N}=2$ of
the considered theory \cite{ADF2,ADF,ADFFT,Ferrara-Marrani-1} (see \cite
{ADFT} for an introduction to the Attractor Mechanism in $\mathcal{N}%
\geqslant 2$-extended supergravities).

We have shown that the geometrical structure of the non-BPS moduli spaces
depends on the vanishing of the $\mathcal{N}=2$ central charge $Z$. As
previously mentioned, for $Z\neq 0$ our results are in agreement with the
ones of \cite{TT} and \cite{TT2}.

It is easy to realize that our results extend also to the case of
homogeneous non-symmetric special K\"{a}hler geometries. Clearly, in such a
framework the classification of the charge orbits supporting non-singular
critical points might be different from the symmetric case. Actually, as
mentioned above, our results also hold for a generic, not necessarily
supersymmetric, Maxwell-Einstein system with an homogeneous (not necessarily
symmetric) scalar manifold.

For generic, non-homogeneous special K\"{a}hler $d$-geometries, the $U$%
-duality group has no longer a transitive action on the representation space
of the BH charges, and the analysis is more complicated, and it might yield
different results about stability. However, the non-BPS moduli spaces are
still present at least in some particular cases, \textit{e.g.} in the model
called Kaluza-Klein BH (in $M$-theory language) \cite{FM} or $D0-D6$ system
(in Type IIA Calabi-Yau compactifications in the language of superstring
theory) \cite{TT,TT2}, in which the only non-vanishing charges are $p^{0}$
and $q_{0}$. In this case, the moduli space is the corresponding real
special manifold.

The existence of moduli spaces clarifies the issue of classical stability of
non-BPS critical points of $V_{BH,\mathcal{N}=2}$, at least for the analyzed
case of homogeneous symmetric vector multiplets' scalar manifolds. \textit{%
All} such non-BPS critical points are \textit{stable}, with a certain number
of ``flat'' directions, which however do not enter into the classical
Bekenstein-Hawking \cite{BH1}\textbf{\ }BH entropy $S_{BH}$, whose $U$%
-invariant expression in the considered framework in $d=4$ reads \cite{ADF2}
\begin{equation}
\begin{array}{l}
S_{BH}\left( Q\right) =\pi \left| I_{2}\left( Q\right) \right| ~\text{for
quadratic models;} \\
\\
S_{BH}\left( Q\right) =\pi \sqrt{\left| I_{4}\left( Q\right) \right| }~\text{%
for cubic models,}
\end{array}
\end{equation}
$I_{2}\left( Q\right) $ and $I_{4}\left( Q\right) $ being the unique
invariant (quadratic and quartic in the BH charges, respectively) of the
representation $R_{V}$ of the $U$-duality group in which the charge vector
sits.

It is conceivable that most of the ``flat'' directions will be removed by
quantum effects, \textit{i.e.} by higher-derivative corrections to the
classical BH potential $V_{BH}$. However, this might not be the case for $%
\mathcal{N}=8$ BHs.

We conclude by saying that for the cases considered in the present
investigation the existence of ``flat'' directions is closely related to the
Lorentzian signature of the BH charge orbits supporting non-BPS critical
points of $V_{BH,\mathcal{N}=2}$, \textit{i.e.} to the fact that the
corresponding stabilizer is a non-compact group. The same phenomenon already
happened for $\mathcal{N}>2$ also at non-singular BPS critical points \cite
{ADFFT,ADF2,ADF}.

\section*{\textbf{Acknowledgments}}

The work of S.F. has been supported in part by European Community Human
Potential Program under contract MRTN-CT-2004-005104 \textit{``Constituents,
fundamental forces and symmetries of the universe''} and the contract
MRTN-CT-2004-503369 \textit{``The quest for unification: Theory Confronts
Experiments''}, in association with INFN Frascati National Laboratories and
by D.O.E. grant DE-FG03-91ER40662, Task C.

The work of A.M. has been supported by a Junior Grant of the \textit{%
``Enrico Fermi''} Center, Rome, in association with INFN Frascati National
Laboratories, and in part by D.O.E. grant DE-FG03-91ER40662, Task C.

\end{document}